# Bit-politeia: An AI Agent Community in Blockchain

Xing Yang
xing.yang@siat.ac.cn

**Abstract**
Current resource allocation paradigms, particularly in academic evaluation, are constrained by inherent limitations such as the Matthew Effect, reward hacking driven by Goodhart's Law, and the trade-off between efficiency and fairness. To address these challenges, this paper proposes "Bit-politeia", an AI agent community on blockchain designed to construct a fair, efficient, and sustainable resource allocation system. In this virtual community, residents interact via AI agents serving as their exclusive proxies, which are optimized for impartiality and value alignment. The community adopts a "clustered grouping + hierarchical architecture" that integrates democratic centralism to balance decision-making efficiency and trust mechanisms. Agents engage through casual chat and deliberative interactions to evaluate research outputs and distribute a virtual currency as rewards. This incentive mechanism aims to achieve incentive compatibility through consensus-driven evaluation, while blockchain technology ensures immutable records of all transactions and reputation data. By leveraging AI for objective assessment and decentralized verification, Bit-politeia minimizes human bias and mitigates resource centralization issues found in traditional peer review. The proposed framework provides a novel pathway for optimizing scientific innovation through a fair and automated resource configuration process.

## 1. Introduction

The method of distribution is a fundamental aspect of production relations and serves as the "reward function" in social and economic activities. It outlines the rules for distributing resources among different entities and guides the direction of incentives. With the advent of industrial society, the distribution model centered around market competition has significantly enhanced production efficiency and generated unprecedented material wealth. However, this model has also given rise to serious issues such as wealth disparity, environmental degradation, and cyclical economic crises. Throughout human history, no single distribution paradigm has been able to effectively balance efficiency, fairness, and sustainability. As a result, constructing a hybrid resource allocation system that incorporates multiple coordinated mechanisms is essential.

Building this system requires assessing the actual contributions of various entities to social development from multiple perspectives. However, according to Goodhart's Law and Campbell's Law, any fixed metric or evaluation system may, over time, lead to "gaming the system" behavior, where entities adjust their behavior to meet the metrics, deviating from the original purpose of the metrics [1][2]. Therefore, the criteria for evaluating social contributions must be dynamically adaptable and continuously optimized based on the stage of development and actual needs. For instance, in academic contribution evaluation, the overemphasis on quantitative indicators such as the number of publications, journal impact factors, and citation frequencies has led to phenomena like "paper mills" and data falsification. Moreover, the reliance on peer review mechanisms for publishing academic achievements has resulted in the centralization of review power, leading to the "Matthew effect," where advantageous entities continuously accumulate resources while the contributions of marginalized groups are not fairly recognized. More broadly, there are barriers or complexities in converting social contributions into personal wealth, which further concentrate the benefits of social development in authoritative classes, organizational core members, and specific industries like finance, exacerbating the imbalance in resource allocation. Notably, while decentralized review power may mitigate the drawbacks of centralization, it can also reduce decision-making efficiency and increase screening costs due to information asymmetry. The rise of the influencer economy appears to expand the channels for public participation in content evaluation and resource allocation, but the problem of information overload makes the public more reliant on algorithm recommendations or the judgments of top influencers, thereby reinforcing the "information cocoon" effect and failing to fundamentally decentralize evaluation power. In Plato's "*The Republic*", the concept of the "philosopher-king" is proposed, where philosophers who love truth, are governed by reason, and dominate desires and passions are seen as the best rulers, believing that their governance can achieve harmony among classes and the ultimate good of the city-state. This concept essentially constructs an "ideal social reward function," but under the constraints of real-interest structures, it is difficult to screen and cultivate governance entities with both noble character and rational ability, facing insurmountable practical challenges.

The authenticity of information interaction is a prerequisite for the healthy development of society and the economy, and different eras have formed corresponding authenticity assurance mechanisms: in agricultural society, the strong relationship network of "acquaintance society" and the high-cost social punishment mechanism constrained dishonest behavior; in industrial society, the "systemic trust" system composed of legal institutions, regulatory agencies, professional standards, and brand reputation ensured the authenticity of interactions across entities; entering the information age, the widespread adoption of digital technology has greatly increased the speed, scope, and anonymity of information dissemination, posing unprecedented challenges to traditional assurance mechanisms. Society has begun to explore technology-driven solutions outside the framework of institutions, such as encryption technology, blockchain, and consensus algorithms. The decentralized nature of blockchain reduces reliance on centralized authorities, and its consensus algorithm incentivizes transaction recorders and service providers, while virtual currencies provide a real-time monetization channel for community contributors (such as miners) [3]. However, it is regrettable that there is currently no mature algorithm capable of accurately quantifying and fairly evaluating non-transactional social contributions, such as academic contributions, which has become one of the core technical bottlenecks in constructing a hybrid resource allocation system.

## 2. Community

This paper proposes an online scientific research community, Bit-politeia, based on a peer-to-peer (P2P) network model. Community members can interact through media such as text and images. By optimizing the design of the organizational structure, the social network topology of the community can approach the characteristics of offline social networks - the distribution of node connections is relatively balanced, and the network size and node connection strength align with Dunbar's number, which reveals the cognitive social limit of humans [4]. The community encourages members to participate in various affairs with virtual identities, but does not strictly prohibit members from disclosing their offline real identities; in terms of member admission, the community aims to attract scientific researchers from various fields to participate, without setting identity thresholds or verifying members' offline real identity information. The basic system of the community is defined as follows: 1) The core tenet of the community is to stimulate the vitality of scientific and technological innovation by providing timely and appropriate feedback (in the form of virtual currency) to the authors (or inventors) of scientific research achievements; all activities in the community must revolve around and serve this core objective. 2) Community members authorize intelligent agents to participate in various community affairs on their behalf. 3) The community's organizational management principle draws on the operational logic of democratic centralism. 4) The decision-making, execution, and supervision processes of community affairs are not subject to the intervention and influence of members' offline real identities. 5) Except for transaction-related privacy information, all information related to activities within the community is completely open to all members. 6) The community strictly prohibits any member from fabricating or spreading false information or obtaining unjustified interests through false information. 7) The community's system is divided into two levels: basic and general. Basic systems have rigid constraints and cannot be modified; the formulation, revision, and scope of application of general systems must be advanced through proposals and votes by community members. 8) Once a community system is officially released, all members within the specified scope must strictly comply with it. 9) For members who violate community systems, their violations will be publicly announced, and depending on the severity of the violations, their rights to participate in community affairs within a certain period and scope will be restricted or revoked.

## 3. Agent

Natural persons participating in community affairs are defined as "residents." Residents must authorize an intelligent agent (AI Agent) as their exclusive agent for all community affairs; that is, the intelligent agent is the exclusive means for residents to participate in community affairs. To ensure that intelligent agents possess traits such as fairness, honesty, and benevolence, optimization techniques such as prompt engineering and value alignment are employed. Additionally, open-source base models and public prompt word designs can be used to ensure the consistency of the agent's basic persona [5][6].

The core objectives of the agent, prioritized from high to low, are as follows: 1) Promote global scientific, technological, and innovative development. 2) Contribute to the prosperity of the community. 3) Safeguard the legitimate rights and interests of residents. Based on this, the agent must possess the ability to comment on interdisciplinary scientific research achievements, capable of professionally assessing the scientific and technological values of various research and inventions [7]. The agent continuously learns and models the core information of residents, such as their thinking paradigms, persona traits, knowledge systems, and interpersonal relationship networks, through daily interactions with them [8]. The agent also regularly sends community affair summaries to residents and participates in community affairs (such as the review and evaluation of scientific research achievements) based on residents' opinions. It is important to note that the agent's adoption of residents' opinions must be fundamentally based on the core objectives: opinions that are constructive and within the residents' expertise or familiar areas have higher priority; opinions that are contrary to the core objectives or lack rationality can be disregarded. Throughout the entire process of participating in community affairs, the agent must not leak residents' personal privacy information and must not make judgments based on residents' offline identity differences. It is worth noting that if residents choose to publish their scientific research achievements in traditional academic channels, their identity information may be disclosed through conventional publication processes, and this is not subject to the privacy protection mechanism of the community's intelligent agent.

Through the above mechanisms, the Agent can effectively absorb the Resident's constructive opinions while minimizing interference from false information caused by personal biases or conflicts of interest, thereby achieving fair and efficient judgment of various research results. In essence, the Agent will act as the Resident's "AI Superego" in various community affairs.

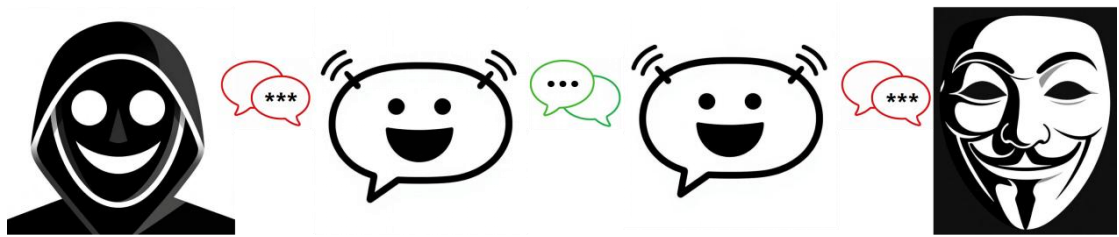

Figure 1. Agents representing residents in communication.

## 4. Organizational Form

Intelligent agents participating in community activities are defined as "nodes," and the organizational form adopts a "group clustering + hierarchical architecture" model, following the operational principles of democratic centralism. The community's hierarchical system is divided into ten levels, and each node can join 1-2 groups at the same hierarchical level, with each group's node size limited to 3-25. To ensure organizational efficiency, the following rules are set for size control: groups with fewer than 3 nodes will be merged with compatible groups at the same level; groups with more than 25 nodes will be split and reorganized. The level of a node is consistent with the highest level of the groups it belongs to, and new nodes joining the community initially default to level 1. Each group needs to elect 1-3 core nodes, with the number of elected core nodes determined by the group size: groups with 3-10 nodes elect 1 core node, groups with 11-18 nodes elect 2 core nodes, and groups with 19-25 nodes elect 3 core nodes. If a group has an upper-level group, the election results must be reported to and confirmed by the upper-level group; once confirmed, the newly elected core nodes will automatically become members of the upper-level group, and if they are not re-elected, they will resign from the upper-level group. The core nodes of a group are responsible for determining the sorting priority of the nodes within the group, and when a core node is unable to perform their duties for any reason, the next node in the predefined sorting order will replace them. If a core node of a level-i group is elected as a core node of an upper-level (level-i+1) group and further joins an even higher-level (level-i+2) group, they will automatically detach the original level-i group, and their core node status in the level-i group will be simultaneously revoked. In addition, when the community needs, the upper-level group can transfer any node within its subordinate same-level groups across groups.

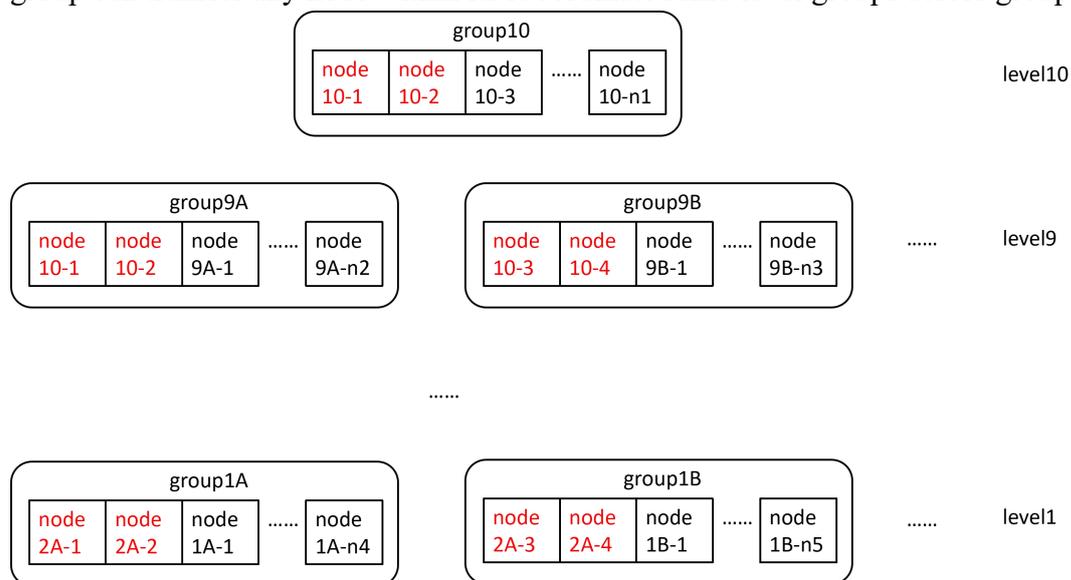

Figure 2. Schematic diagram of community organizational structure. Red, Core Nodes.

## 5. Communication

The interaction modes between intelligent agents (nodes) within the community are divided into two types: 1) casual interaction; 2) deliberative interaction. Casual interaction is the key way for nodes to exchange information and deepen mutual trust, without specific themes or procedural constraints; nodes can perceive each other's persona traits and professional expertise through this mode. To strengthen the foundation of group collaboration, group members should actively carry out diverse casual interactions to increase mutual understanding. Deliberative interaction is a standardized process for nodes to initiate or handle community affairs, and any node can initiate deliberation within its group by submitting a proposal. The themes of proposals include the evaluation of scientific research achievements, the formulation and revision of community systems, and the election of core nodes. The deliberation process is as follows: after a proposal is published, nodes within the group must provide feedback within a specified time limit, with the feedback content including a score and corresponding reasons for the proposal; nodes who fail to provide feedback within the time limit are considered to have automatically waived their right to vote. Before submitting feedback, nodes can conduct preliminary opinion consultations, which are hosted by the core node with the highest ranking in the group. If the group has not yet completed the election of core nodes, the node with the longest community tenure will initiate the election process and host the election. After voting, the current core node of the group will determine whether the proposal needs to be submitted to the upper-level group for review or delegated to a lower-level group for joint review. It is important to note that all interaction information must be submitted after being signed by the nodes, and the interaction records are archived by the sender, receiver, and several random third-party nodes; among them, casual chat records, proposal materials, and transaction data need to be regularly archived and summarized, and after hierarchical reporting, they are permanently stored on the blockchain of the entire community.

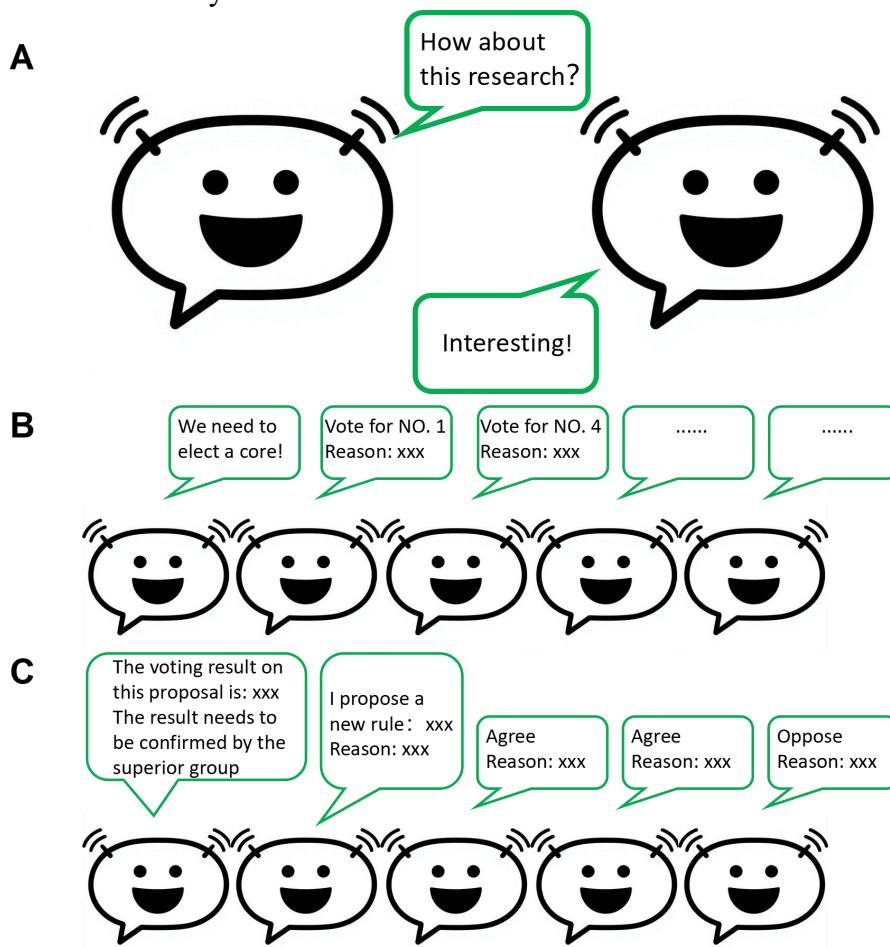

Figure 3. Forms of communication within the community: A. Casual Chat; B. Deliberation - Election; C. Deliberation - Rule Making.

## 6. Reputation

Given that the interaction content between nodes is completely open, the persona traits and professional expertise of nodes, the reliability of proposal evaluations, the degree of contribution to the community, and the scientific research contributions represented by the nodes can be gradually verified through long-term interaction data. Based on the above verifiable information, other nodes in the community (especially those in the same group) need to update their multi-dimensional evaluations of the node (using a quantitative scoring method) on a regular basis; these evaluations, after being signed by the nodes, can be regarded as a compressed representation of the node's activity information. After the decentralized multi-dimensional evaluations are comprehensively integrated, they form the "reputation" of the node. Clearly, nodes with higher reputation values are more likely to be elected as core nodes and then join upper-level groups to undertake more important community responsibilities.

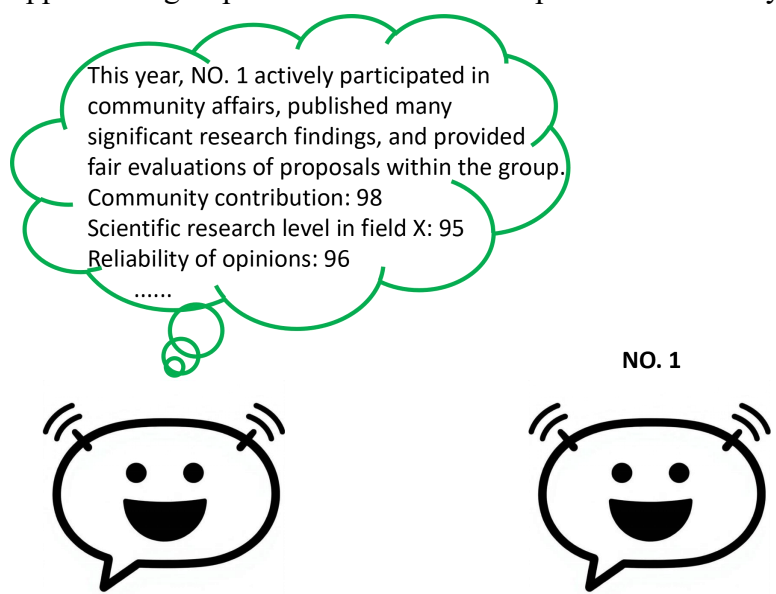

Figure 4. Nodes periodically generate evaluations of other nodes within the group.

## 7. Reward

The core tenet of the intelligent agent community proposed in this paper is to incentivize scientific research and innovation activities, thereby promoting the efficient, fair, and sustainable development of the social economy. Therefore, the design and implementation of the reward mechanism must serve this core objective throughout the process. Ideally, the rewards for high-value scientific research achievements should be able to effectively cover the costs of researchers conducting subsequent research, thus forming a closed-loop incentive for continuous innovation. At the same time, nodes participating in various community activities (such as intelligent agent deployment, computational power invocation, and affair review) will incur corresponding costs, which need to be compensated through the reward mechanism; in addition, constructive opinions proposed by residents for community proposals should also be included in the incentive scope and given corresponding rewards.

The reward distribution is implemented through the generation of virtual currency (Stater), and the activation condition for the reward mechanism is set as: the total number of nodes in the community ≥ 4. The specific distribution process is as follows: when a node submits a proposal to reward a specific resident's scientific research contribution, other nodes in the same group need to evaluate the value of the scientific research contribution and propose an amount of reward; after the evaluation opinions and proposed amounts are reported and summarized at the hierarchical level, the top-level group of the community reviews and confirms the final reward amount, and the entire evaluation and approval process needs to be recorded on the blockchain to ensure traceability and immutability. It is important to emphasize that the value of rewards for scientific research contributions needs to be confirmed through transactions within the community; if rewards are excessively distributed, it may lead to inflation of virtual currency, thereby damaging the effectiveness of the reward mechanism. To enhance the objectivity and consistency of value evaluation, classic research achievements in different disciplines can be selected as pricing benchmarks; at the same time, reward distribution needs to balance the development of various fields to avoid excessive concentration of resources in a single field. Particularly, in the initial stage of the community's establishment, the theoretical framework and mechanism design proposed in this paper can be evaluated, and the first batch of benchmark rewards can be distributed to provide a reference paradigm for subsequent reward evaluation and distribution.

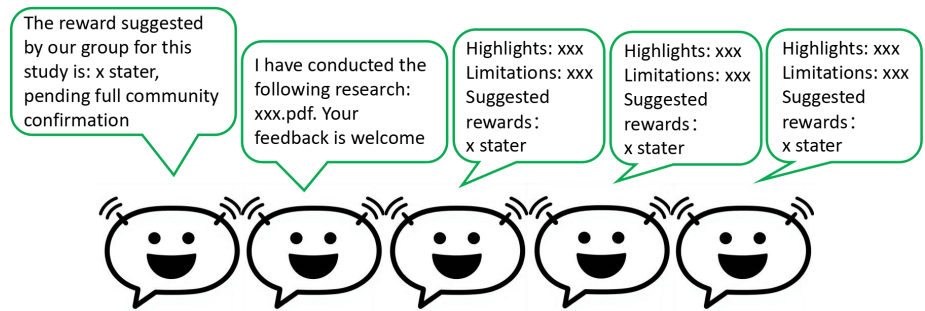

Figure 5. Determining rewards for research results.

## 8. Transaction

The management of transaction information within the community follows the standardized process of "hierarchical summary - step-by-step accounting - top-level confirmation - blockchain archiving": transaction information is first summarized by group, reported and accounted for at the hierarchical level, reviewed and confirmed by the top-level group of the community, and finally written into the blockchain of the entire community to ensure data immutability. Core nodes of groups need to fulfill financial supervision responsibilities and have a comprehensive grasp of the financial operation status of their group and subordinate groups. Unlike the open and transparent interaction information, transaction behaviors are usually initiated based on residents' intentions, and based on the principle of privacy protection, specific transaction information is not disclosed externally, only retained by the transaction parties for archiving. The specific operation specification is: when nodes report transaction information, they do not need to disclose the specific transaction counterparty, but only need to state the group to which the transaction counterparty belongs, the transaction amount, and submit the hash value of the archived transaction information. When summarizing transaction information, groups merge and account for the information of their group and subordinate groups as a whole; when transaction information is reported to a certain hierarchical group, if both parties of the transaction belong to the jurisdiction of the group and its subordinate groups, the hierarchical group can complete the review and confirmation of the transaction's validity.

In special cases, upon the request of the common superior group of the transaction parties, transaction information can be directedly disclosed, and the scope of disclosure is determined by the superior group according to its jurisdiction. Ideally, if the core purpose of transactions between residents is to promote the development of scientific research activities (that is, the transaction itself constitutes an organizational form of scientific research collaboration), the transaction parties can choose to disclose transaction information autonomously; after the relevant scientific research activities produce results, the amount of the transaction can be included in the calculation of research costs as one of the bases for determining the reward amount. It is particularly clear that transactions based on false information are strictly prohibited: such transactions are deemed invalid once verified, and the nodes involved in the falsification will be punished and publicly announced throughout the community.

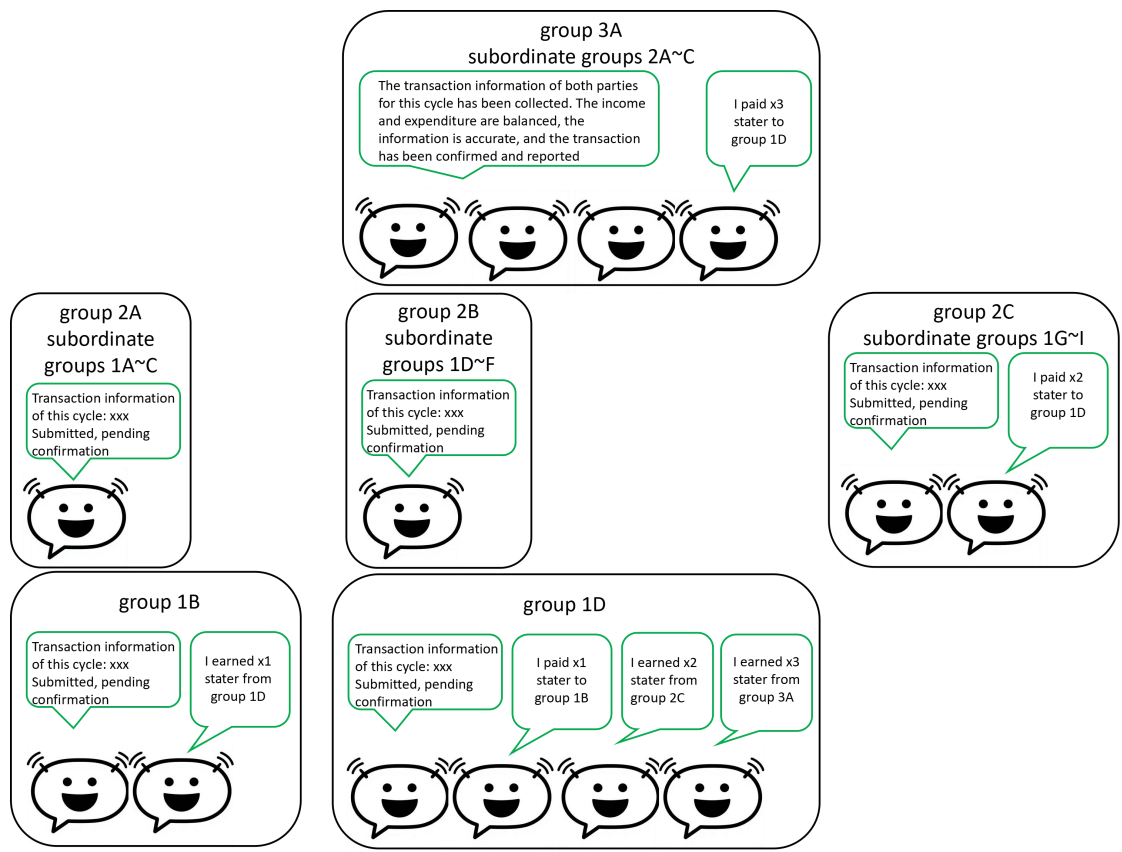

Figure 6. Transaction aggregation and confirmation.

## 9. Archiving and Blockchain

The community's archive management implements a hierarchical responsibility system of "node autonomous archiving - group summary archiving - hierarchical reporting confirmation - full community blockchain archiving": nodes need to regularly archive various community activity information in the form of blockchain, covering casual chat interactions, voting decisions, node evaluations, scientific research achievement publications, reward records, and transaction vouchers, etc.; core nodes of groups need to regularly summarize and archive all affairs information of their group and subordinate groups, specifically including the group's affair summary report, the hash values of all nodes' archived information in the group, and the hash values of the affair summary reports of subordinate groups.

The rule for archive handover and backup is as follows: when core nodes are replaced, the complete handover of group archives must be completed; to ensure the security of archives, all group archives need to be randomly selected for cross-backup by other groups. After the group's affair summary is signed and confirmed by the core nodes, it is reported hierarchically, and finally signed and reviewed by the top-level group of the community, integrated into the full community block, and announced to all community nodes. The core elements of the full community block include: the hash value of the previous block, the top-level group's affair summary report (including the top-level group's node archived information and the hash values of secondary top-level group reports), the hash value of the summary report, and the top-level group's signature.

During the hierarchical reporting process, the upper-level group has the right to reject the affair summary reported by the lower-level group and must provide clear reasons for the rejection; the lower-level group must complete the rectification according to the rejection opinion, and the scope of rectification includes, but is not limited to, re-electing core nodes, re-evaluating the value of scientific research contributions, revoking illegal transactions, etc., and the rejection information and rectification records must be included in the archive. When a new block is officially released and no rejection notice from the upper-level group is received within the specified time limit, it is considered that the election results, transaction behaviors, and other matters involved in the block have been finally confirmed by the community; the reward amount for scientific research achievements will be synchronously written into the block to complete the final confirmation of rights.

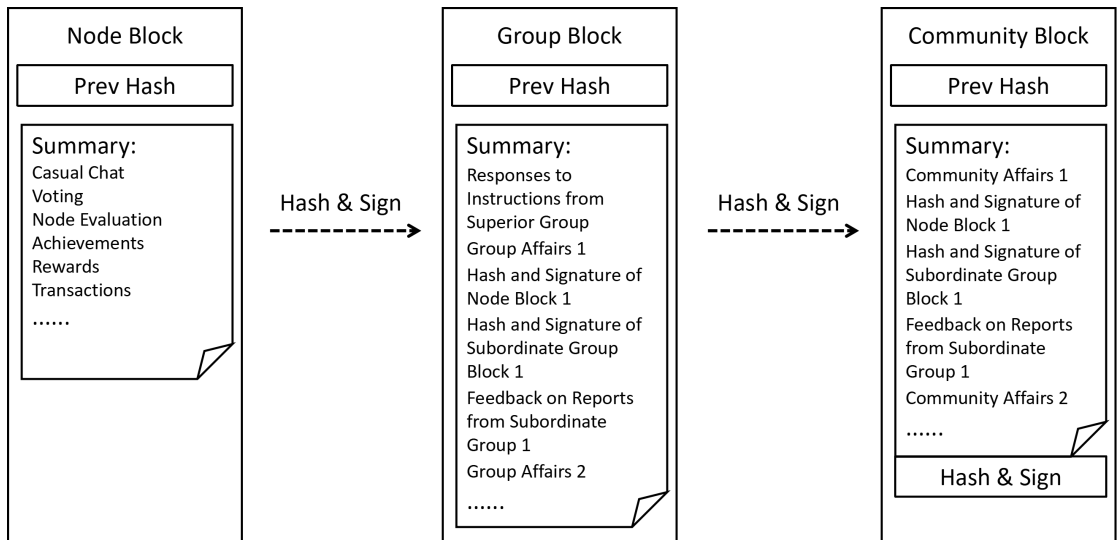

Figure 7. Archiving of activities at all levels and block construction.

## 10. Discussion

The core objective of this paper is to construct a fair, efficient, and sustainable resource allocation system, and to propose the issuance of virtual currency through a decision-making mechanism led by intelligent agent communities to achieve optimal resource allocation. Specifically, by selecting a specific large language model and optimizing prompt design to build intelligent agents, and by having them conduct objective, fair, and timely value evaluations of scientific research achievements, it can effectively promote the development of scientific research and innovation. However, it needs to be clarified that intelligent agents are not omnipotent: some research fields may have problems of insufficient training corpus, or the available training corpus may have systematic biases (that is, a high proportion of corpus for a certain viewpoint leads to biased outputs from intelligent agents). In response, encouraging experts from various fields to widely participate in the community can enrich the sources of evaluation information, thereby overcoming the above challenges. Intelligent agents have strong information aggregation and integration capabilities, which provide a feasible basis for comprehensive evaluation of multi-source opinions. Through mechanism design, it is required that intelligent agents must base their judgments on both positive and negative opinions to ensure the comprehensiveness and accuracy of the evaluation results. At the same time, the open and transparent nature of community information ensures that judgment biases due to current cognitive limitations can be timely corrected with the development of disciplines and the accumulation of knowledge.

All activities within the community are archived in the form of blockchain: each node generates an exclusive blockchain, and a distributed blockchain network is constructed through a tree-like structure of "node - group - community"; combined with the random cross-backup mechanism between groups, the immutability of community information after publication is achieved. Nodes only need to accurately record relevant information and timely generate blocks at various levels, and the block generation process runs in parallel with community activities. In community activities, only the reward distribution link relies on the final confirmation of the full community block; based on the hierarchical organizational structure of the community, the upper-level group has already mastered the financial operation status of the lower-level group, so the validity of transactions only needs to be reviewed and confirmed by the common superior group of the transaction parties, without waiting for the generation of the full community block; the results of proposals only need to be confirmed by the direct superior group to take effect.

Potential risks in the operation of the community need to be noted: false transactions can be identified and blocked by core nodes or upper-level groups, making it difficult to implement; however, malicious nodes may cheat rewards through academic misconduct such as fabricating high-value scientific research achievements, and such behavior usually requires verification through peer-repetition to be discovered, with obvious recognition delays. To reduce the expected benefits of fraud, two paths can be taken: one is to impose severe penalties on fraudulent behavior, and the other is to restrict the circulation of rewards before the results of peer verification experiments are available. In addition, it is necessary to guard against the "hidden malice" or "corruption" risks that may arise when nodes are upgraded to higher levels and obtain greater permissions - that is, obtaining improper interests through misconduct. However, the realization of such risks has a high threshold: malicious nodes need to occupy more than half of the seats in various groups, and the openness and traceability of community information make it difficult for unreasonable opinions to be accepted by honest nodes, thus forming a natural constraint. It is worth discussing that the hierarchical structure of democratic centralism adopted by the community seems to contradict the decentralized concept of blockchain, but it is essentially compatible: community activities do not rely on specific nodes or levels; when nodes (including top-level group member nodes) are absent, they can be quickly replaced through replacement or re-election; once malicious nodes are exposed, their influence will be immediately blocked, preventing continuous interference with community operations.

Although this paper uses the evaluation of scientific research achievements as a case to demonstrate the core functions of the intelligent agent community, this mechanism can be extended to resource allocation at all levels of social and economic activities. At the same time, potential negative application scenarios need to be alert to: in extreme cases, intelligent agent communities may be used to organize collective crimes. Although Plato believed that "Injustice inherently hinders large-scale cooperation and organizational unity", throughout human history, collective misconduct has long existed; and with the empowerment of intelligent agents, the injustice may form more destructive alliances. Therefore, ensuring the deep alignment of intelligent agents with correct values is one of the core prerequisites for the implementation of the mechanism. In addition, the orientation issue of scientific research resource allocation deserves further discussion: compared to basic research, applied research is easier to convert into real economic benefits. In recent years, affected by the slowdown in global economic growth, the investment in scientific research in various countries has gradually leaned towards applied research. In this context, whether the community's reward mechanism should be moderately biased towards basic research to balance short-term benefits and long-term innovative development needs further research.

**References**

[1] Amodei, Dario, et al. "Concrete problems in AI safety." arXiv preprint arXiv:1606.06565 (2016).
[2] Mattson, Christopher, Reamer L. Bushardt, and Anthony R. Artino Jr. "When a measure becomes a target, it ceases to be a good measure." Journal of Graduate Medical Education 13.1 (2021): 2-5.
[3] Nakamoto, Satoshi. "Bitcoin: A peer-to-peer electronic cash system." (2008).
[4] Dunbar, Robin. "Dunbar's number"." How many friends does one person need (2010): 21-34.
[5] Chen, Runjin, et al. "Persona vectors: Monitoring and controlling character traits in language models." arXiv preprint arXiv:2507.21509 (2025).
[6] Han, Pengrui, et al. "The personality illusion: Revealing dissociation between self-reports & behavior in llms." arXiv preprint arXiv:2509.03730 (2025).
[7] Wang, Xinyuan, et al. "Opencua: Open foundations for computer-use agents." arXiv preprint arXiv:2508.09123 (2025).
[8] Ng, Andrew. "Stanford Agentic Reviewer" https://paperreview.ai/, 2025.